\def\bm#1{\mbox{\boldmath $#1$}}

\documentclass[a4paper,12pt]{article} 
\usepackage{geometry}
\usepackage{authblk}

\usepackage{amsmath}
\usepackage{amsmath,amssymb}
\usepackage[utf8]{inputenc}
\usepackage{empheq}
\usepackage{graphicx}
\usepackage{subcaption}
\begin{document}

\title{Efficient electron transfer in quantum dot chains controlled by a cubic detuning profile via shortcuts to adiabaticity}

\author[1]{María E. Rus}
\author[1]{Rodolfo H. Romero }
\author[1]{ Sergio S. Gomez }
\affil[1]{
 Instituto de Modelado e Innovaci\'on Tecnol\'ogica, Universidad Nacional del Nordeste, CONICET, Facultad de Ciencias Exactas y Naturales y Agrimensura, Avenida Libertad 5400, W3404AAS Corrientes, Argentina.}
\maketitle

%\date{\today}

%\pacs{73.21.La}{Quantum dots}
%\pacs{02.30.Yy}{Control theory}
%\pacs{03.67.-a}{Quantum information}

%\begin{abstract}
\abstract{
Long-distance fast and precise transfer of charge in semiconductor nanostructures is one of the goals for scalable electronic devices. We study theoretically the control of shuttling of an electron along a linear chain of semiconductor electrostatically-defined quantum dots by an electric field pulse with nonlinear time-dependent profile. We show that this essential nonlinearity along with shortcuts to adiabaticity techniques speed up the electron transfer with high fidelity, while still holding great robustness under spin-flip interactions and inhomogeneities in the couplings of the chain.
A given fidelity can be set experimentally by controlling the maximum sweep energy and duration of the control pulse.
}

\section{Introduction}

Fast and coherent manipulation of states of charge and spin in nanostructures is envisioned as a key requirement for feasible scalable platforms for quantum based technologies, such as quantum information processing \cite{Bruss-Leuchs19, Bauerle18}. Gate-controlled semiconductor quantum dots (QD) \cite{Kloeffel13, Hanson07} are one of the most investigated experimental realizations for that goal, due to their potential scalability with well-stablished fabrication techniques of semiconductor industry along with improvements in the control of impurities and coherence times \cite{Zwanenburg13, Gonzalez21}.

In particular, linear arrays of semiconductor QDs has been advocated as a gate architecture for scalable quantum devices \cite{Zajac18, Kandel21}. 
This systems are interesting candidates to the realisation of quantum computing, with experimental implementations where both charge and spin can be controlled in GaAs, Si or InSb based quantum dots nanostructures \cite{Fujita19, Flentje17, Bertrand16}. These dots are created in the interface of an heterostructure, like GaAS/AlGaAs, where a two-dimensional electron gas is confined, and the couplings between them can be experimentally tuned to specific values by means of electrostatic gates. 
Also, recently two dimensional structures were implemented, showing that scalability in this type of systems is achievable in short term\cite{Mortemousque21}.  
Several other physical systems are also currently being explored as feasible implementations of quantum devices \cite{Zhou16, Kjaergaard20, Alexandre21}.

The present work focus on the development of a reliable protocol for rapidly transfer an electron along this system. 
%\subsection{Protocols}
There are several proposals to accomplish this task, e.g. Spatial Adiabatic Passage \cite{SAP16} and Coherent Transfer by Adiabatic passage \cite{Platero13, Platero18, Platero19, Petta20}, consisting on techniques based on the Stimulated Raman Adiabatic Passage (STIRAP) \cite{STIRAP1, STIRAP2}. Another variant,  also based in the STIRAP strategy, uses the concept of User Defined Passage (UDP) \cite{Niu22}, where inverse engineering is used to define a specific control strategy on a three-level system. A different approach is the strategy of physical unitary transformations (PUT) in conjunction to shortcuts to adiabaticity (STA) \cite{Berry09, Ibanez12, Torrontegui19}, designed to speed up the transfer produced by a uniform electric field along an adiabatic state by introducing a counterdiabatic potential unitary transformed to provide feasible control fields \cite{Gomez19, Rus21}.

%%%%%%%%%%%%%%%%%%%%%%%%%%%%%%%%%%%%%%%
%\subsection{Field profile}
The fidelity of the protocol can be strongly affected by the spin-orbit interaction or imperfections in in the system setting \cite{Kloeffel13, Khomitsky12, Srinivasa13, Schreiber11}.
The use of nonlinear time control profiles was proven to increase the fidelity of the protocol even without the inclusion of STA terms\cite{Vitanov99, Kam20, Dou18, Torosov11}. One desirable features of the field is to accelerate the dynamics when the levels are well separated ($t=\pm \infty$), and therefore with low probability of transition, and to slow its pace as the levels approach each other ($t=0$) to minimize their mixing. Such a strategy should already provide a speedup as compared to a process of the same fidelity controlled by linear profile field. Higher improvement could be achieved by using a shortcuts to adiabaticity technique. 
%%%%%%%%%%%%%%%%%%%%%%%%%%%%%%

In this work, we derive analytical expressions for the probability of shuttling of an electron with defined spin projection between the ends of a linear chain of quantum dots. The dynamics is driven by a uniform electric field with an arbitrary time-dependent profile and the counterdiabatic Hamiltonian calculated for speeding up the process. Then, the PUT protocol is applied to obtain approximate analytical expressions for the particular case of a cubic profile as compared to the linear Landau-Zener one. Finally, dephasing effects of spin-orbit interaction and tunneling coupling inhomogeneities are studied numerically. The calculations show the robustness of the protocol as measured by the fidelity to the target state.

\begin{figure}
\begin{center}
 \includegraphics[width=11.cm]{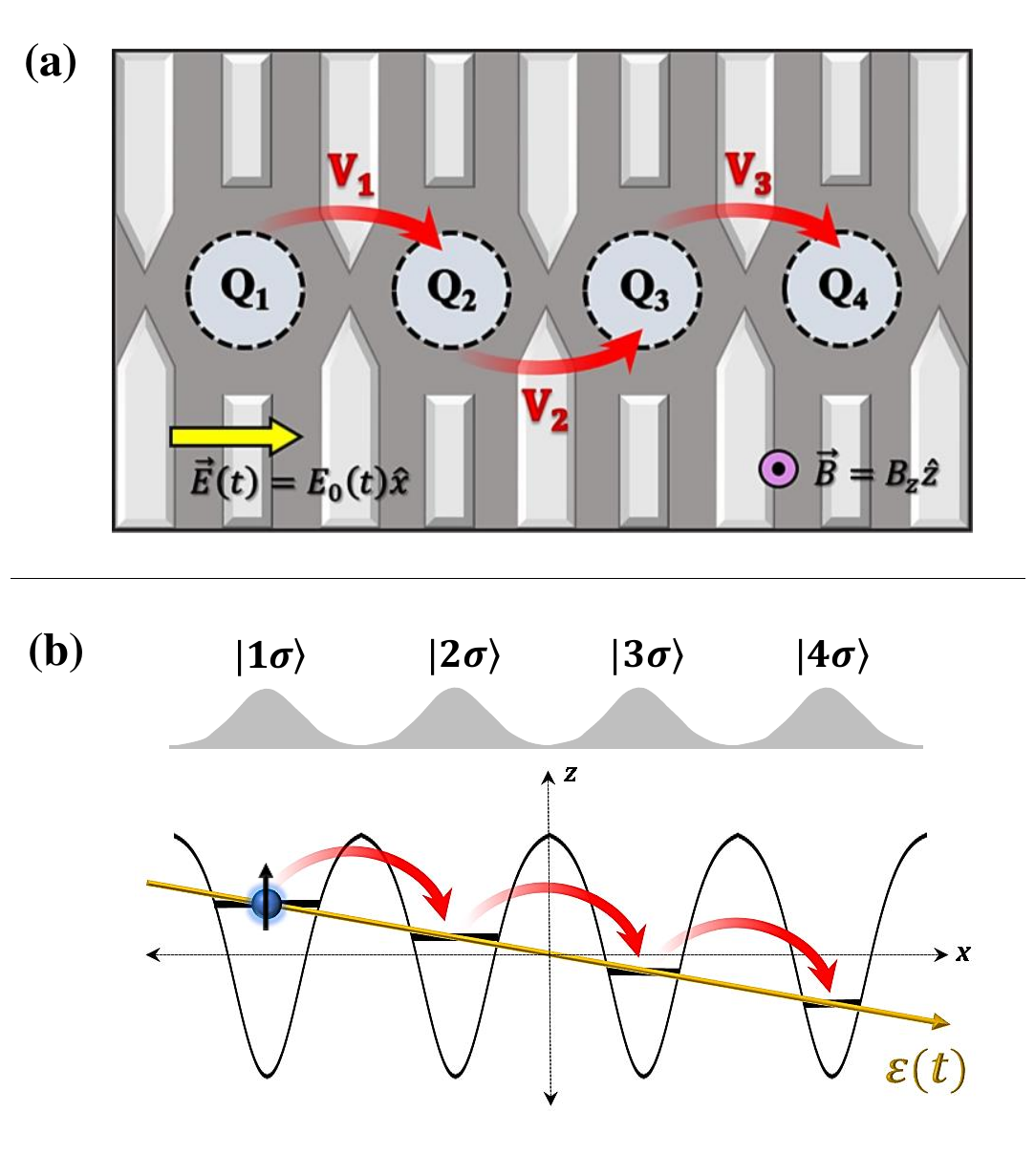} \\ 
\caption{(a) Scheme of the device for a chain of four coupled quantum dots. The shapes in light-gray represents the gates that allow to experimentally define the value of the couplings $V_i$. (b) Scheme of the electrostatic potential along the chain produced by the electrostatic gates. The electron is initially placed at the first site and a time-dependent electric field is applied in the direction of the chain to transfer it; a constant magnetic field is also present to decouple the electron spin states. 
\label{fig1}}
\end{center}
\end{figure}
%=================================
\section{Theory}
We consider a linear chain of $N$ tunnel-coupled adjacent QDs placed at $x_k=k d$ ($k=1,\ldots,N$), equally spaced a distance $d$, and subject to a uniform electric field $E(t)$, schematically depicted in Fig. \ref{fig1}. 
We chose a $S$-spin model of QD chain, with nonuniform tunneling couplings $V_k$, amenable to analytical treatment \cite{Rus21}, described by a Hamiltonian $H^{(0)}_{\sigma}=H_t+H_E(t)$ including the tunneling and electric potential contributions for given spin projections $\sigma=\uparrow,\downarrow$
%=========================
\begin{eqnarray}
H_t &=& \sum_{k,\sigma=\uparrow,\downarrow} V_k |k\sigma\rangle \langle k+1,\sigma| + {\rm H.c.}, \nonumber \\
H_E(t) &=& \sum_{k,\sigma=\uparrow,\downarrow} \varepsilon_k(t) |k\sigma\rangle \langle k\sigma| %= -e E(t) \sum_{k,\sigma=\uparrow,\downarrow} x_k |k\sigma\rangle \langle k\sigma|,
\label{Hamiltonian}
\end{eqnarray}
%========================
with $ V_k = -V\sqrt{(N-k)k}/2 \label{eq:hoppk}$, $(k=1,\ldots,N-1)$, and $\varepsilon_k(t)=k\varepsilon(t)=-e x_k E(t)$. Similar particular cases of multistate models have been studied in the context of LZ transitions \cite{Sinitsyn04, Sinitsyn16}.

A static magnetic field $B_z$ perpendicular to the chain is added to split the electron spin states 
$
H_{\rm Z,\sigma} = 
%\Delta \sum_{k} \sigma | {\rm k}\sigma \rangle \langle {\rm k}\sigma| = 
\Delta \left(|k\uparrow \rangle \langle k\uparrow|-| k\downarrow \rangle \langle k\downarrow|\right),  \label{dqd-ham-zeeman}
$
where $\Delta=g\mu_B B_z$ is the Zeeman splitting, $g$  Land\'e factor of the material and $\mu_B$ is the Bohr magneton. 
Therefore, the complete Hamiltonian is $  H^{(0)}= \sum_{\sigma} \left( H^{(0)}_{\sigma} + H_{\rm Z,\sigma} \right) $.

Formally, $H^{(0)}_{\sigma}$ can be written as $H^{(0)}_{\sigma}= -\bm{B}_0\cdot \bm{S}$, the Hamiltonian for a Zeeman interaction of a spin $S=(N-1)/2$ with a time-dependent magnetic field $\bm{B}_{0} =\left(-V ,0, \epsilon(t)\right)$, i.e.
 %------------------------
\begin{equation}
H^{(0)}_{\sigma} =  \varepsilon(t) S_z -  V S_x,
\label{eq:HSC}
\end{equation}
%------------------------
This interpretation is useful for choosing the physical unitary transformations (PUT) leading to experimentally feasible Hamiltonians \cite{Rus21, Ibanez12}.
%%%%%%%%%%%%%%%%%%%%%%%%%%%%%%%%%%%%%%%%%
\subsection{PUT strategy}
Shortcuts to adiabaticity techniques speed up the state evolution, $i\hbar \partial_t|\psi\rangle = H^{(0)} |\psi\rangle$, along an adiabatic state from  an initial state $| \psi_j(t_i) \rangle$ to the target state $| \psi_f(t_f) \rangle$ \cite{Berry09, Torrontegui19}. That is achieved by adding a {\em counterdiabatic} correction term $H_{CD}(t)$  suppressing transitions to nonadiabatic levels. For $H^{(0)}_\sigma$, Eq. (\ref{eq:HSC}), it becomes \cite{Rus21}
%
%-------------------
\begin{equation}
H_{{\rm CD},\sigma} =  -i \hbar\frac{ \partial_t\varepsilon(t) V}{(  \varepsilon(t)^2 + V^2 )} S_y 
\label{HCD}
%h^{(1)} &=& -\frac{\xi_{LZ}V}{((  t/\tau_{LZ} )^2 + 1 )}.
\end{equation}
%------------------------
%-------------------------
Applying a further unitary transformation $U=\exp{(-i\Phi S_z)}$ to $H_\sigma(t)=H_{0,\sigma} +H_{{\rm CD},\sigma}$ in order to suppresses non-physical imaginary couplings between the QDs (i.e., between $S_z$ eigenstates), the Hamiltonian becomes
%----------------------------------
\begin{eqnarray}\label{eq:HamSTA}
H^{(0)}_{{\rm PUT},\sigma} = \left(\varepsilon(t) -\hbar \dot{\Phi}(t) \right)S_z -   V(t) S_x,
\end{eqnarray}
%------------------------------------
where $V(t)$ and $\Phi(t)$ are peak-like pulses having a duration of the order of nanoseconds for typical QD nanostructures, implementable via control of the gate potentials of the devices \cite{Gomez19}.
%%%%%%%%%%%%%%%%%%%%%%%%%%%%%%%%%%%%%%%
\subsection{Electric field profiles}
The most studied profile is an electric field ramp $\varepsilon^{\rm LZ}(t)=\lambda t$ increasing linearly on time, due to its relation to the solutions of the two-level LZ problem and its experimental feasibility. Nevertheless, using an accelerated nonlinear field profile could improve the speed and efficiency for charge transfer. Nonlinear level crossing dynamics have studied analitically in two-level systems \cite{Vitanov99, Kam21}.
In particular, we  consider here a cubic detuning electric field $\varepsilon^{\rm C}(t) =  \varepsilon^{\rm C}_0 \left( t/\tau_E\right)^{3}$,  with amplitude $\varepsilon^{\rm C}_0$ and time scale $\tau_E=\hbar/V$. It has a vanishing time derivative at the avoided crossing, unlike the linear LZ profile $\varepsilon^{\rm LZ}(t) =  \varepsilon^{\rm LZ}_0 \left(t/\tau_E\right)$ widely studied both numerically and theoretically \cite{Sinitsyn04,Sinitsyn16,Sinitsyn18,Fuxiang17,Chernyak18,Chernyak21}.
We aim to study the effect of nonlinearity in cubic level crossing in the $N$-state chain. 

The protocol for the electron transfer is designed to operate in a short time window $(-T,T)$. For sake of comparison between cubic and linear profiles, we require both to reach the same maximum energy $\varepsilon^{\rm C}(T)=\varepsilon^{\rm LZ}(T)=\varepsilon_{\rm max}$ at the end of the evolution. The {\em mean energy rate} $\lambda_m = \varepsilon_{\rm max}/T$ allows one to relate the amplitudes of both profiles as $\varepsilon^{\rm C}_0=(\varepsilon^{\rm LZ}_0)^3/\varepsilon_{\rm max}^2 = (\hbar/V T)^2 \varepsilon^{\rm LZ}_0$. 
It should be noted that the cubic and linear amplitudes are of different orders for a given sweep energy $\varepsilon_{\rm max}$.
For example, for a typical value of $\varepsilon_{max}= 1$ meV and a LZ amplitude around $\varepsilon^{LZ}_0 = 0.2$ meV, the cubic amplitude is $\varepsilon^{C}_0 =  0.008 \varepsilon^{LZ}_0$. 

%------------------ RHR -----------------
The {\em infidelity} $I$ of a state evolution (as opposed to the fidelity $F$) is the measure of the failure of the dynamics of a state $\psi(t)$, evolving during the interval $(t_i,t_f)$, to reach a given target state $\Psi$ at $t_f$, namely, $I=1-F=1-|\langle \Psi|\psi(t_f)\rangle|^2$. Aiming to quantum information applications, we impose to the electron transfer protocol the requirement that $I \leq 10^{-4}$, a common threshold for fault-tolerant quantum computing \cite{Bruss-Leuchs19}.
%%%%%%%%%%%%%%%%%%%%%%%%%%%%%%%%%%%%%%%%%%%%%%%%%%%%%%%%
\section{Results}
\subsection{Accuracy of linear and cubic crossings}
We start by analyzing the accuracy of transfer under linear and cubic crossing for linear chains (without the speedup PUT protocol) as related to the corresponding transitions in two-level systems.

%------------ RHR ---------------------
The transfer of an electron from the first (at $t=t_i$) to the last site of the chain (at $t=t_f$), driven by a linear crossing of speed $\lambda$, amounts to start and finish the process at the ground state of the chain. For a TLS, the probability of such a process is $1-P_{\rm LZ}$, where $P_{LZ}=\exp(-\pi V^2/2\hbar\lambda)$ is the Landau-Zener transition probability. For $\lambda\ll V^2/\hbar$ the evolution is approximately adiabatic with infidelity $I=P_{LZ}\ll 1$. 
For a chain of $N$ sites the infidelity of the $|1\rangle\rightarrow|N\rangle$ transfer becomes $I= 1-(1-P_{LZ})^{N-1}$, which for nearly adiabatic evolution (small $\lambda$), becomes $I \approx (N-1)P_{LZ}$.

\begin{figure}[b!]
\begin{center}
\includegraphics[width=10.0cm]{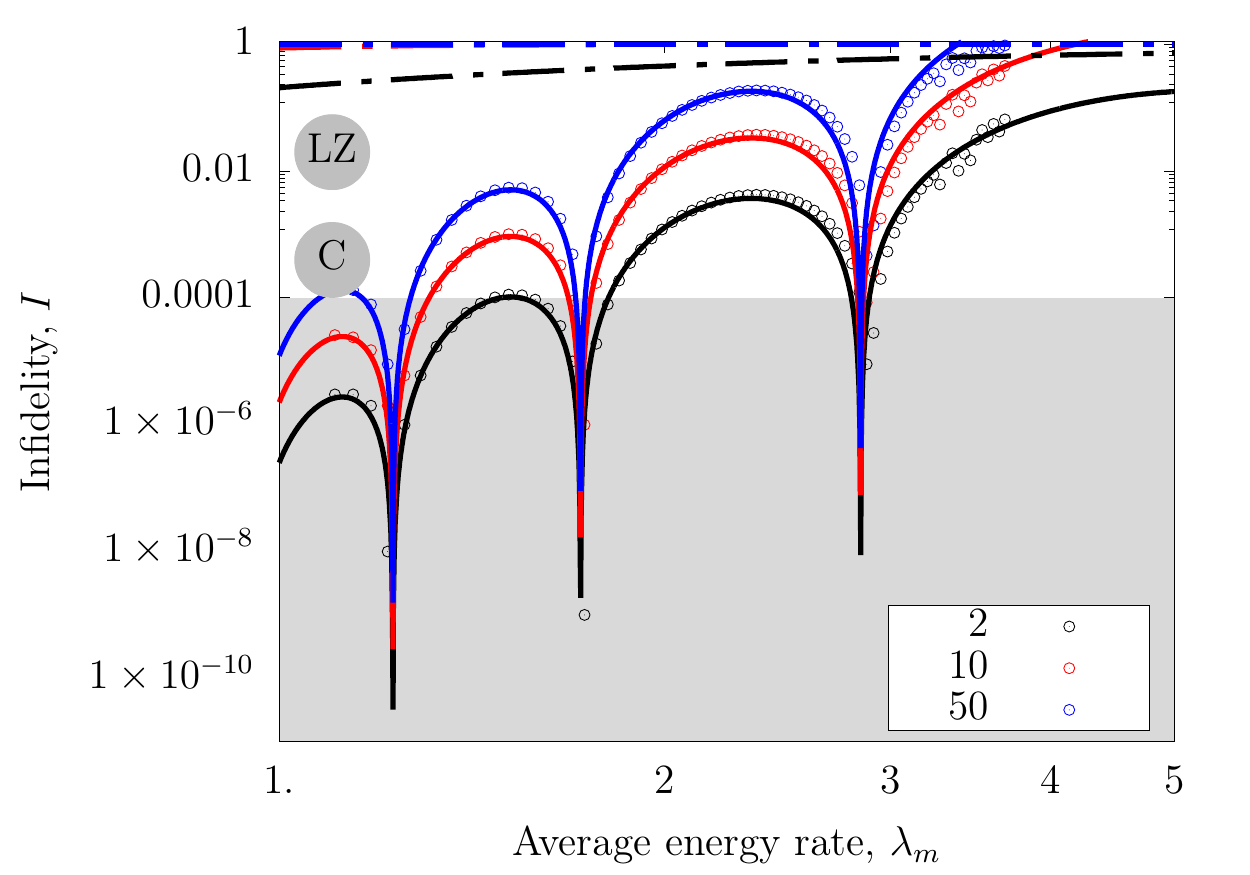} \\ 
\caption{The calculated infidelity $I=1-P^{\rm av}_{N\uparrow}$ for the transfer of the electron state along a linear chain of $N$ sites, driven by linear (LZ) and cubic (C) time-dependent profiles, plotted as a function of the mean energy rate $\lambda_m$, for three sizes $N=2,10,50$ sites. Linear driving (dashed lines) are orders of magnitude less accurate than cubic ones represented as symbols {\large $\circ$} (numerical calculations) and solid lines for the extrapolation proportional to $N-1$ discussed in the text.\label{fig2}}
\end{center}
\end{figure}

For the cubic level crossing, an approximate expression $P^{C}_{\rm TLS,\sigma}$ for the probability of the transitionless dynamics of a TLS was given in \cite{Vitanov99}. We numerically calculated the probability of electron transfer for chains of various lengths $N$, and assume them to extrapolate linearly with the chain length, $I^{\rm C} = (N-1)P^{C}_{\rm TLS,\sigma}$. Fig. \ref{fig2}, shows the infidelity $I$ as a function of the mean level rate $\lambda_m$ for three chain lengths, $N=2, 10$ and 50, for linear (LZ) and cubic (C) level crossings. Numerical calculations are shown with open circles. Remarkably, the proposed extrapolation, represented in solid lines, fits to them with good precision. 
%%%%%%%%%%%%%%%%%%%%%%%%%%%%%%%%
%%%%%%%%%%%%%%%%%%%%%%%%%%%%%%%%%%
The gray shaded bottom region of Fig. \ref{fig2} represent the useful region of work for accurate dynamical processes ($I\leq 10^{-4}$).
As $\lambda_m$ increases, neither the linear nor cubic profiles reaches this sought fidelity. 

Therefore, although a clear improvement is achieved with the cubic crossing over the linear one, it is not enough for high accuracy. Further improvement is introduced by adding the counterdiabatic potential given by the PUT protocol, Eq. (\ref{eq:HamSTA}).
%%%%%%%%%%%%%%%%%%%%%%%%%%%%%%%%%%%%%%%%
%\section{Results}
\subsection{Analytical model for the probability of electron shuttling}
%%%%%%%%%%%%%%%%%%%%%%%%%%%%%%%%%%%
%\begin{figure}[h]
\begin{figure*}
\begin{center}
$\begin{array}{cc}
\includegraphics[width=8.0cm]{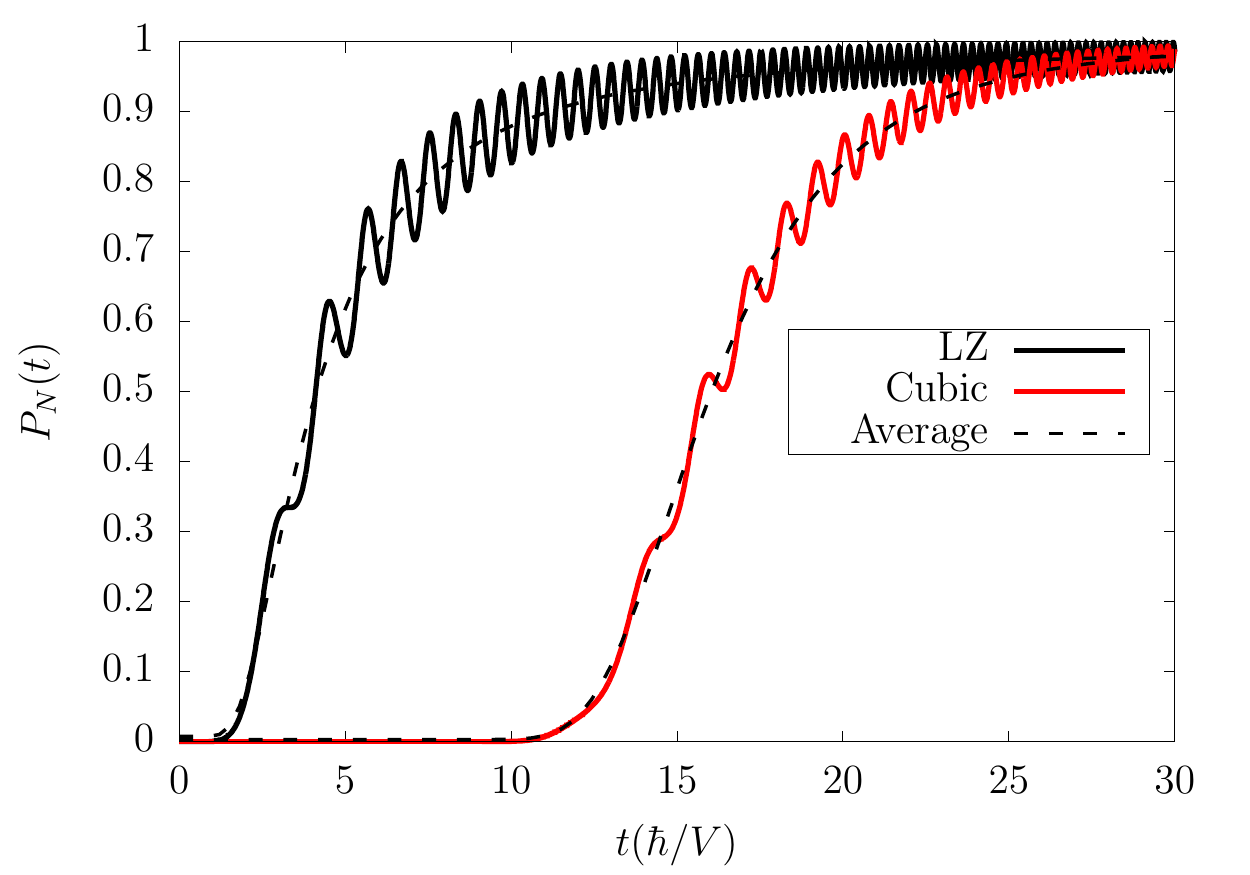}  &
\includegraphics[width=8.0cm]{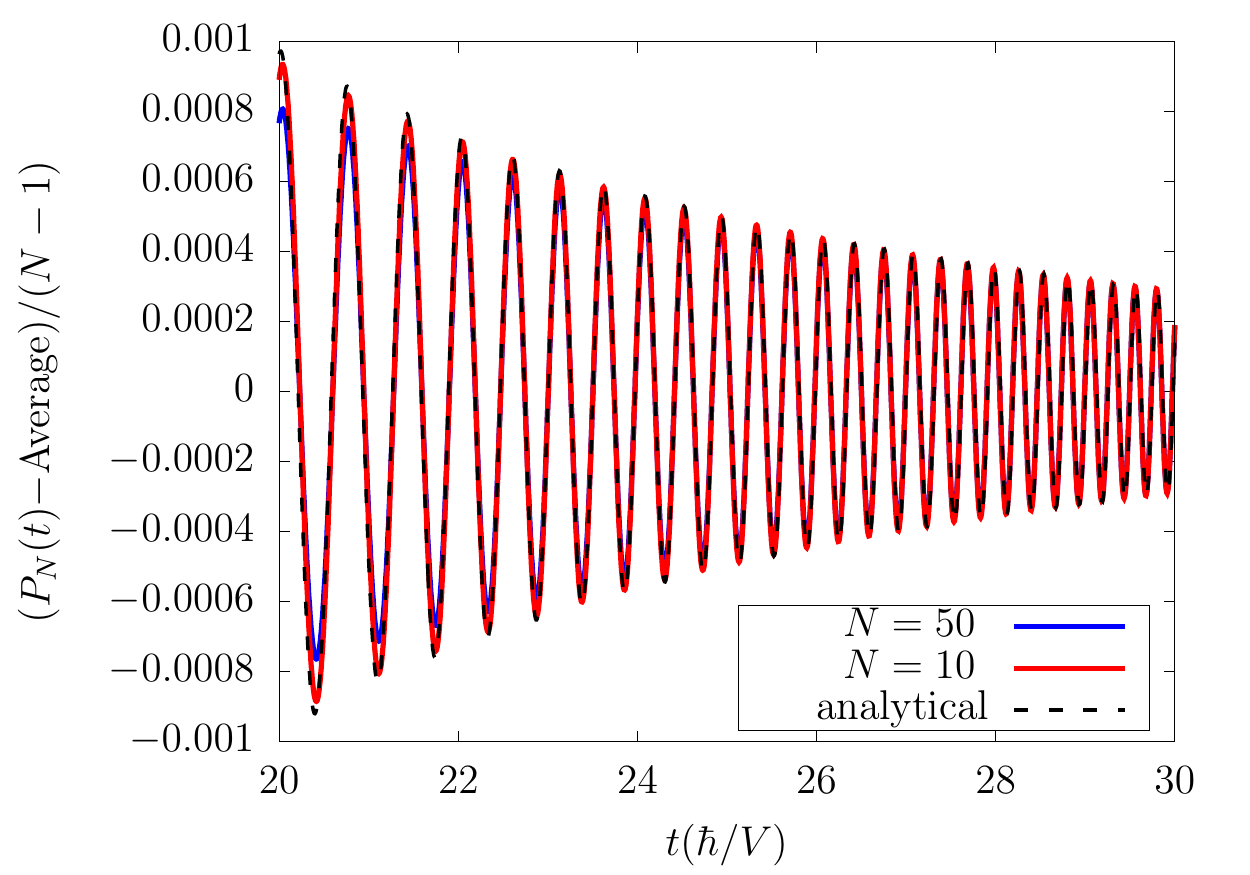} \\ 
\end{array}$
\caption{(a) Time evolution of the probability of shuttling the electron from the first to the last site $N$ of the chain calculated numerically with the PUT Hamiltonian, Eq.( \ref{eq:HamSTA}), for $N=50$, for the linear (solid black) and cubic (solid red) profiles. Mean energy rates are $\lambda_m = 1$ for linear crossing, and $\lambda_m = 0.65$ (in units of $V^2/\hbar$) for cubic crossing. The average probability $P_N^{\rm av}(t)$, Eq. (\ref{eq:Model}), for the model described in the text is shown in dashed lines. 
(b) The difference between the numerically calculated probability $P_N$ and the average analytical probability, Eq. (\ref{eq:Model}) for cubic crossing, rescaled by $1/(N-1)$, in two chains of 10 and 50 sites. The transfer is driven within the interval $(-40,40)$, in units of $\hbar/V$.
\label{fig:propevol} }
\end{center}
\end{figure*}

For the $S$-spin chain considered here, an analytic expression for the target probability $P_N(t)$ can be obtained with initial and final conditions at the end sites of the chain along an infinitely long adiabatic process evolving from $|\psi(-\infty)\rangle = |1\rangle$ to $|\psi(\infty)\rangle = |N\rangle$. 
Recalling that the state $|1\rangle$ of the chain corresponds to the eigenstate $|S\rangle$ of $S_z$, Eq. (\ref{eq:HSC}), the instantaneous adiabatic ground state is $R_y(\theta)|S\rangle$, where $R_y(\theta)$ is a rotation of an angle $\theta(t)=\arccos{(\varepsilon(t)/V)}$ around $y$-axis. 

By applying the PUT transformation, Eq. (\ref{eq:HamSTA}), to eliminate complex couplings from the Hamiltonian, an approximate expression for the probability $P^{(0)}_N(t)$ at long times $t$ is obtained from the transformed states, 
%==========================
\begin{eqnarray} \label{eq:PNinf}
P^{(0)}_{N}(t) =  \frac{1}{2^{N-1}}\left(1+\frac{\varepsilon(t)}{\sqrt{V^2+\varepsilon(t)^2}}\right)^{N-1}.
\end{eqnarray}
%=========================

Nevertheless, we are interested in the probability of shuttling the electron within a finite range $(-T,T)$ as required for the fast control of real systems. Then, taking the initial state as $|\psi(-T)\rangle = |1\rangle = \sum_n c_n|\psi_n(-\infty)\rangle$, i.e., as a linear combination of adiabatic states at $t=-\infty$, heavily weighted on $|\psi_0(-\infty)$, we get
%-----------------------------------------
\begin{equation}
P_{N}(t) = P^{\rm av}_{N}(t) + \frac{(N-1)V^2}{2\varepsilon_T \varepsilon(t)} \cos{\alpha(t)},    
\label{eq:Model}
\end{equation}
%-----------------------------------------
%where
%-----------------------------------------
\begin{equation}
P^{\rm av}_{N}(t) = P^{(0)}_{N}(T)P^{(0)}_{N}(t) +(1-P^{(0)}_{N}(T))(1-P^{(0)}_{N}(t))    
\label{Prob average}
\end{equation}
%------------------------------------------
is $P^{\rm av}_{N}(t)$ a monotonic function of time and the second term of (\ref{eq:Model}) oscillates with a time-dependent frequency $\hbar \alpha(t) = \int_{-T}^{t} \left( E_{1}(t')-E_{0}(t')\right) dt'$, given by the accumulated dynamical phase between the ground and the lowest excited adiabatic states. $P^{\rm av}_{N}(t)=\overline{P_{N}(t)}$ is the time average of the probability
within a period of these oscillations.
It should be noted that $P^{\rm av}_{N}(t)\approx P^{(0)}_{N}(t)$ [Eq. (\ref{eq:PNinf})] if $P^{(0)}_{N}(T)\approx 1$.
Eq. (\ref{eq:Model}) is valid for both linear and  cubic crossings and for arbitrary length $N$.\\

Figure \ref{fig:propevol}, shows the results of numerical calculations of $P_N(t)$ obtained from the Hamiltonian $H^{(0)}_{{\rm PUT},\sigma}$, Eq. (\ref{eq:HamSTA}), as compared to the probability $P_{N}(t)$ of the analytical model, Eq. (\ref{eq:Model}). 
Figure \ref{fig:propevol}a, depicts numerically calculated $P_N(t)$ along a $N=50$ chain, for both linear (solid black lines) and cubic (solid red lines) crossings.  
Mean level rates are $\lambda_m= V^2/\hbar$ (linear), and $\lambda_m=0.65 V^2/\hbar$ (cubic), chosen to separate both curves. The average probability of the model, $P^{\rm av}_{N}(t)$, is shown in dashed lines. 

Figure \ref{fig:propevol}b compares the oscillating part of the probability in more detail. It shows $(P^{\rm num}_N(t)-P^{\rm av}_{N}(t))/(N-1)$, for cubic crossing along chains of two lengths ($N=10$, 50). $P^{\rm av}_{N}(t)$ is subtracted to show the oscillations and a rescaling by a factor $1/(N-1)$ is introduced to compare different chain lengths. The overlap of both curves shows that the analytical expression  [Eq. (\ref{eq:Model})] describes very accurately the PUT dynamic of the electron for moderate and large times.

From the model, the oscillations have a time dependent frequency $\varepsilon(t)/2$ (linear) and $\varepsilon(t)/4$ (cubic); the amplitude of the oscillations decreases as $1/\varepsilon(t)$ for both linear and cubic crossings. 

\subsection{Protocol}
Now we can state a prescription for a fast and accurate control of the process. 

Given a fixed high fidelity $F=1-I$, a pulse $\varepsilon(t)$ is applied, that reaches its maximum value $\varepsilon_{\rm max}$ at the end of the interval $(-T_I,T_I)$, with  $T_I\gg \hbar/V$. 

The fidelity of the process can be estimated as  $F=1-I=P_{N}^{\rm av}(T_I)$. 
Using Eq. (\ref{Prob average}) at $t=T_I$, and solving for $P_N^{(0)}(T_I)$, we get $P_N^{(0)}(T_I)=1-I/2$, where we have assumed $I\ll 1$.

On the other hand, from Eq. (\ref{eq:PNinf}), $P_N^{(0)}(T_I)$ can be expressed in terms of the system parameters $N$, $V$ and $\varepsilon_{\rm max}$. Assuming $V\ll \varepsilon_{\rm max}$, results
\begin{equation}
P_N^{(0)}(T_I)=1-\frac{N-1}{4}\left(\frac{V}{\varepsilon_{\rm max}}\right)^2.    
\end{equation}
%----------------------------
Equating both expressions for $P_N^{(0)}(T_I)$ and solving for $\varepsilon_{\rm max}/V$, we get
%----------------------------
\begin{equation}
    \frac{\varepsilon_{\rm max}}{V}=\sqrt{\frac{N-1}{2I}}.
    \label{eq:epsvsI}
\end{equation}
%----------------------------
For instance, to get $I=10^{-3}$, in a chain of $N\sim 20$ and $V\sim 10-100$ $\mu$eV , a pulse $\varepsilon_{\rm max}\sim 1-10$ meV should be applied during a time $T_I\gg \hbar/V \sim 0.1$ ns, within the reach of current experiments.

%%%%%%%%%%%%%%%%%%%%%%%%%%%%%%%%%%%%%%%%%%%%%%%%%%%%%%%%%
%\subsection{Decoherence due to SOI and couplings defects}
Two sources of decoherence affect the fidelity of state transfer considered in this work: the spin-orbit interaction (SOI) and the imperfections in the control couplings. 
%====================================================
\begin{figure*}[h]
\begin{center}
$\begin{array}{cc}
{\rm Linear \ crossing} & {\rm Cubic \ crossing} \\
\includegraphics[width=7.5cm]{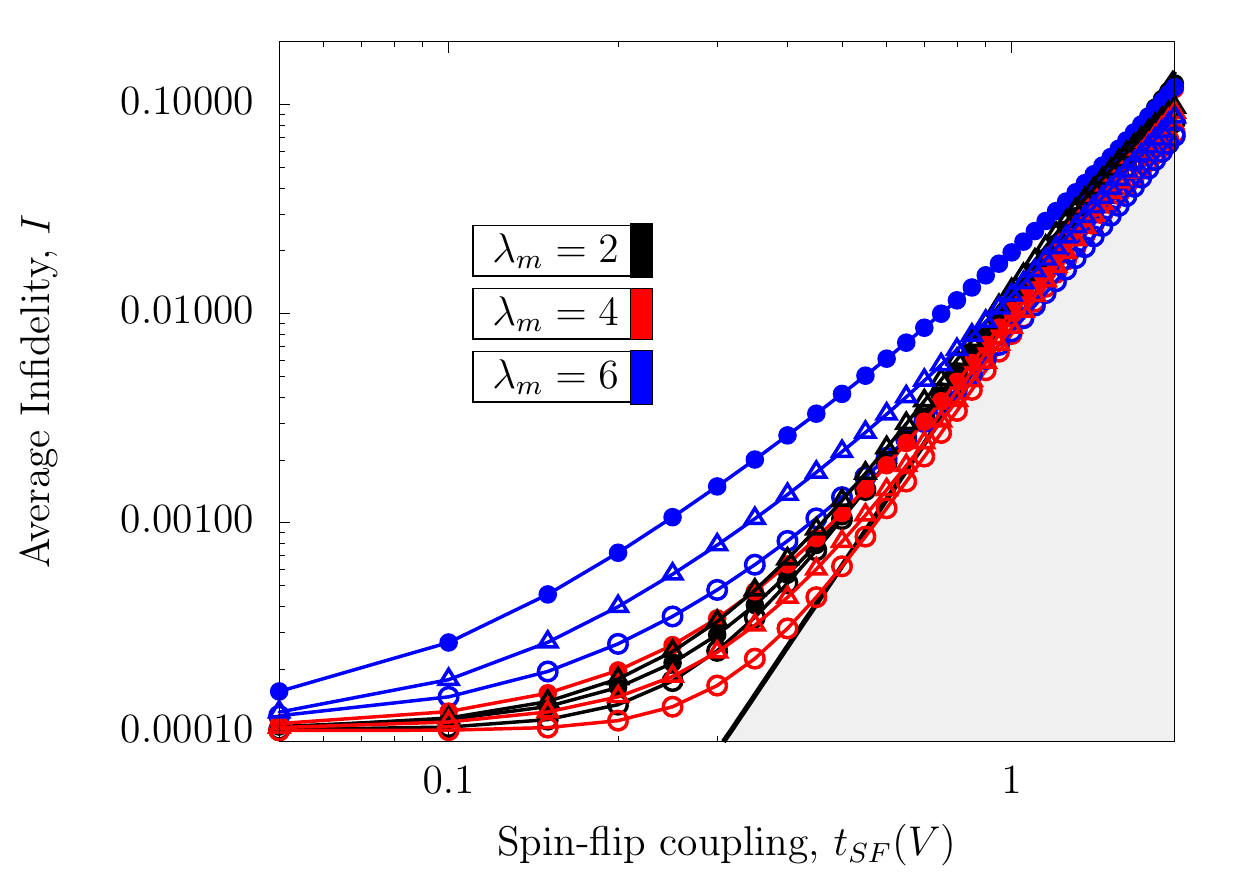} & \includegraphics[width=7.5cm]{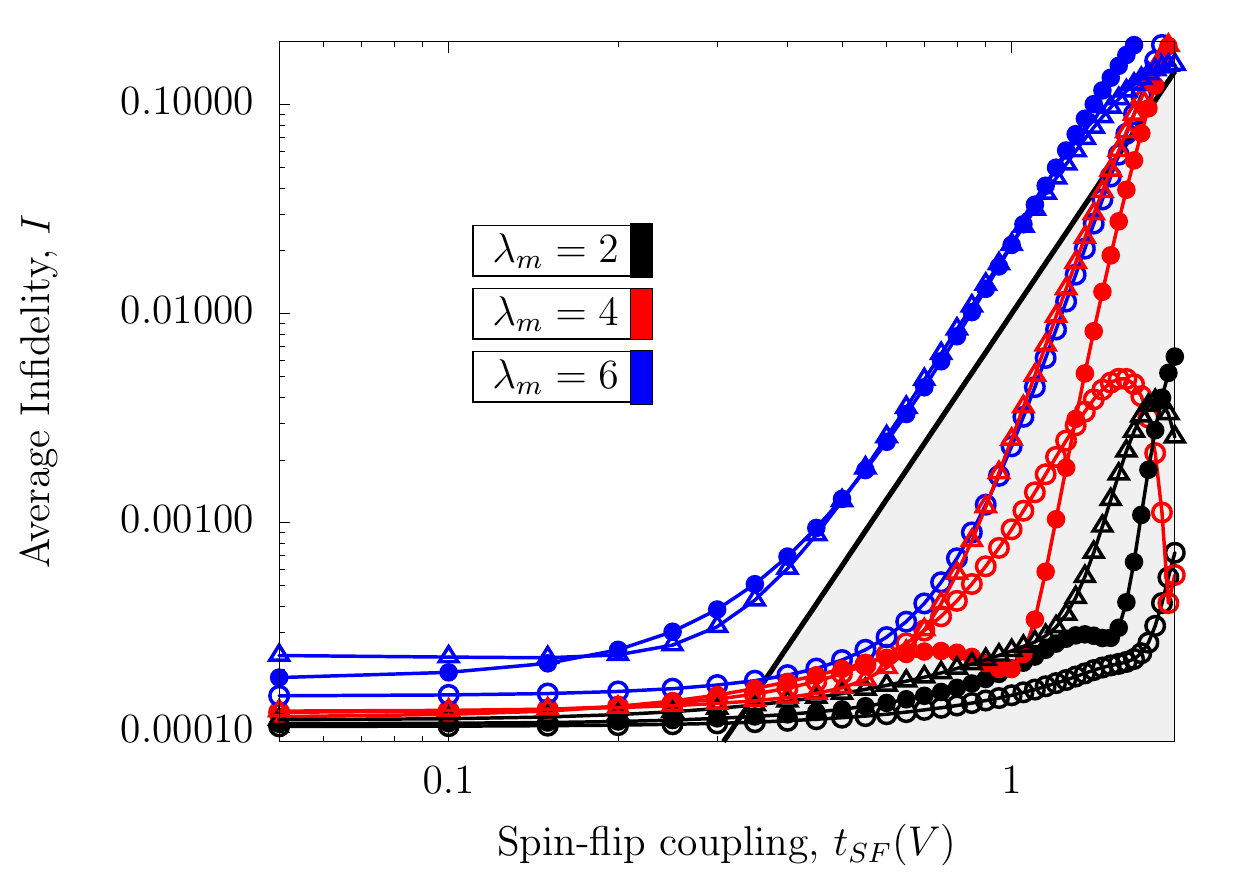} \\
\end{array} $ 
\caption{Numerical calculations of $I=1-P^{\rm av}_{N\uparrow}$ as a function of the spin-flip coupling amplitude $t_{\rm SF}$, for linear (left)  and cubic (right) control profiles on three chain lengths, $N=4$ ({\Large$\bullet$}), $N=6$ ($\triangle$) and $N=10$ ({\Large $\circ$}). Three values of the mean energy rate are shown: $\lambda_m=2$ (black),  4 (red) and  6 (blue) in units $V^2/\hbar$. The control time $T_I$ was taken to achieve $I=10^{-4}$ with no spin-flip, as explained in the text. The slanted straight line of the shaded area represents the $t_{\rm SF}^4$ dependence.
Spin-flip effects are smaller for the cubic profile for all values of $t_{\rm SF}$ used, providing a more accurate transfer to the target state than the linear one.\label{fig SOI}}
\end{center}
\end{figure*}
%====================================================
%%%%%%%%%%%%%%%%%%%%%%%%%%%%%%%%%%%%%%%%%%%%%%%%%%%%%
\subsection{Decoherence due to SOI}
The SOI allows spin-flip (SF) transitions during the electron transfer, so that there is a leakage to opposite spin projection states and any information encoded in the spin projection cannot be preserved. It originates in the Rashba or the  Dresselhaus mechanisms, and in the one-dimensional tight binding approximation, is given by
%==============================
\begin{eqnarray}
H_{\rm SF}=-t_{\rm SF}\sum_{n} |n+1,\uparrow \rangle \langle n \downarrow | -|n+1,\downarrow \rangle \langle n \uparrow |  + {\rm H.c.},  
\end{eqnarray}

%===============================
where $t_{\rm SF}$ stands for the SF coupling strength between neighbors sites on the chain. 
In typical semiconductors, $t_{\rm SF}$ can be a small fraction of the spin conserving coupling $V$, as in GaAs, or as large as twice the spin-conserving coupling $V$, as in InSb or InAs, or even larger \cite{Gomez19}. 
SOI induces lost of fidelity on the dynamics of the analytical model presented in this work due it does not include SOI corrections to (i) adiabatic states nor (ii) to the counteradiabatic potential $H_{\rm CD,\sigma}$. We briefly discuss them in the following.

(i) Adiabatic states are corrected through first order : 

$| \Psi_n(t_{\rm SF}) \rangle \approx | \Psi_n(0), \uparrow \rangle + (t_{\rm SF}/\Delta E) | \Psi_{\perp}(0), \downarrow \rangle ,$ 
introducing a leakage to $\downarrow$-spin orthogonal adiabatic state $| \Psi_{\perp}(0)$, with $\Delta E \sim \varepsilon(t)\rightarrow \infty$ at large $T$. Therefore, the correction to the norm gives an additional $I \sim {\cal O}(t_{\rm SF}^2)$, decreasing with time.

(ii) Counterdiabatic Hamiltonian, Eq. (\ref{HCD}), is also corrected both due changes in $(dH_0/dt)_{mn}$ and energy differences $E_m-E_n$, through $\delta H_{\rm CD}(t_{\rm SF}) \sim t_{\rm SF}^2$, which modifies the dynamical evolution to the adiabatic target state by an error $I \sim {\cal O}(t^2_{\rm SF}) + {\cal O}(t^4_{\rm SF})$

Figure \ref{fig SOI} shows the results of numerical calculations of the infidelity $I$ with the PUT Hamiltonian as a function of the strength $t_{\rm SF}$ for three chain lengths ($N=4$, 6 and 10) and three mean level rates ($\lambda_m=2$, 4 and 6 $V^2/\hbar$). Calculations were performed with the linear (left panel) and cubic (right panel) level crossing profiles.

For the linear crossing (left panel), the error  $I$ to reach the target state increases with $t_{\rm SF}$ for a given chain length $N$. The scaling $I\displaystyle \propto t_{\rm SF}^4$, indicated by the straight line of the shaded area, describes the qualitative behaviour of $I$ for the linear profile at high $t_{\rm SF}$ for all sizes $N$ and rates $\lambda_m$.
For the cubic crossing (right panel), the dependence $I(t_{\rm SF})$ presents a more complex behaviour. Nevertheless, it can be seen that most of the curves are within the shaded area, under the line $I\displaystyle \propto t_{\rm SF}^4$. The fastest level crossing ($\lambda_m=6$) is the most sensitive to SOI following a $I\displaystyle \propto t_{\rm SF}^4$ behaviour, even at moderate strengths. 
In general, the cubic profile is less sensitive to SOI than the linear one. A more sophisticated model than the one studied here would be needed to explain the general behaviour of the cubic crossing.

%%%%%%%%%%%%%%%%%%%%%%%%%%%%%%%%%%%%%%%%%%%5
\subsection{Decoherence due to couplings defects}
Imperfect realization of the couplings of the particular chains can also hinder the accurate transfer of the state. To assess their influence on the control, we performed numerical calculations with defects on the couplings $V_n$ between the sites of the chain, modeled as $V_n \rightarrow (1+ \xi x_n)V_n$, where $x_n$ is a random variable with uniform distribution in the range $(-1/2,1/2)$, and $\xi$ the maximum allowed amplitude of the defects.

\begin{figure}[h]
\begin{center} 
\includegraphics[width=10.0cm]{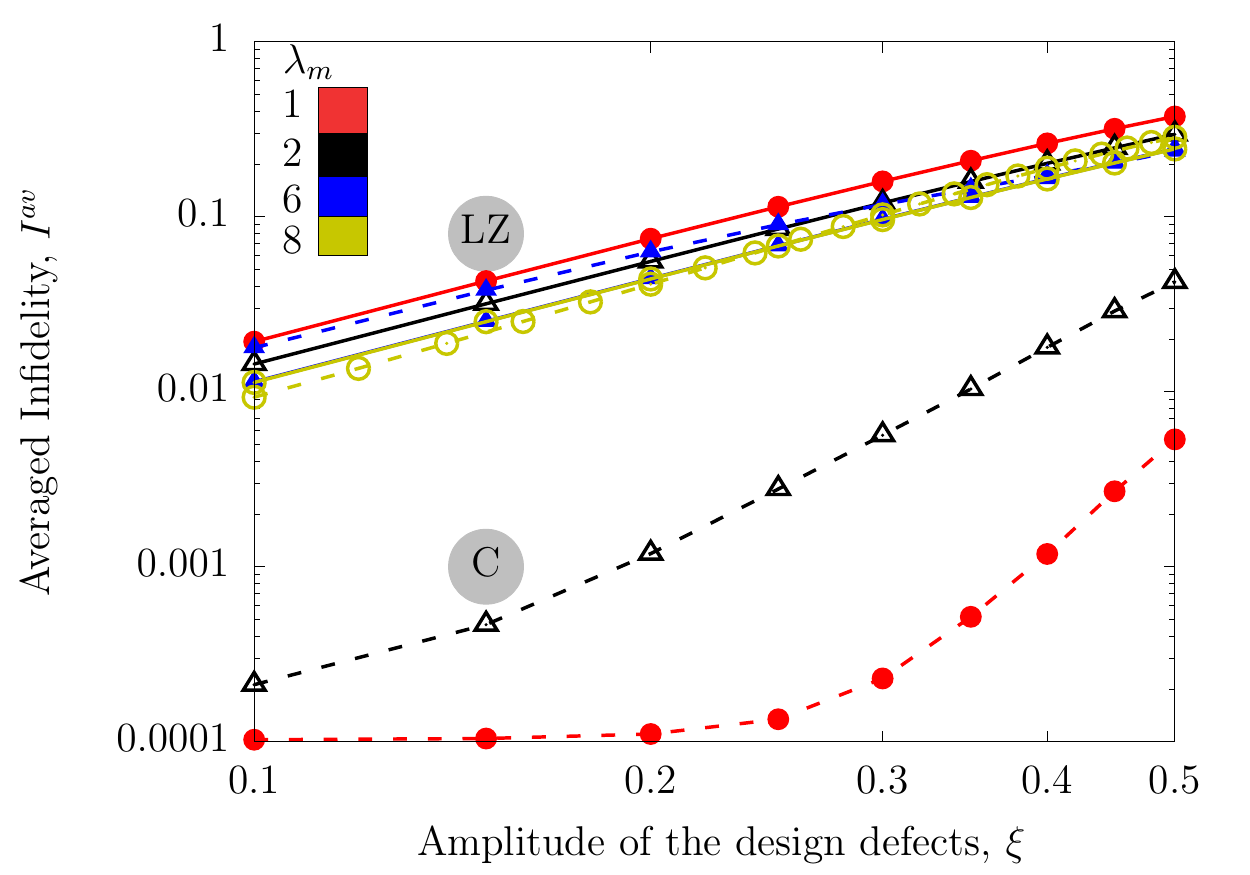}
\caption{  $I=1-P^{\rm av}_{N\uparrow}$ as a function of the amplitude of the design defects $\xi$, for three mean energy rates: $\lambda_m=1$ ({\Large $\bullet$}), $\lambda_m=2$ ({\small $\triangle$}) , 
$\lambda_m=6$ ($\blacktriangle$) and $ \lambda_m=8$ ({\Large $\circ$}) in units of $V^2/\hbar$, for both linear (LZ, solid lines) and cubic (C, dashed lines) profiles for a chain of $N$=10 sites. The interval of control $T_I$ is defined in the text.}\label{fig7}
\end{center}
\end{figure}

Fig. \ref{fig7} shows the result of the mean infidelity $I^{\rm av}$ as a function of the defect amplitude $\xi$, calculated by averaging over $10^3$ different realizations of the state dynamics by sampling $V_n$, for mean energy rates $\lambda=1$, 2, 6 and 8 $V^2/\hbar$. 
The processes with linear driving are more sensitive than cubic ones to decoherence induced by defects in couplings, roughly depending on their amplitudes as $\xi^2$ for all energy rates $\lambda_m$. The cubic drivings are weakly dependent on $\xi$ for slow speed ($\lambda_m \lesssim 2)$, while for the fastest crossings ($\lambda_m \gtrsim 6$) have the same order of accuracy and $\xi$-dependence as the linear one, without a definite monotonic dependence on $\lambda_m$.

%=============================================================
%================================================================

\section{Conclusions}
We studied the transfer of an electron along a linear chain of quantum dots, driven by linear and cubic time-dependent electric pulses of short duration. 
The cubic crossing becomes more accurate than the linear one for the same maximum energy and duration of the pulses. 
For high precision requirements, we performed numerical calculations with the Hamiltonian $H_{\rm PUT}$  including additional shortcuts to adiabaticity terms which speeds up the process. The particular model of the chain allows us to develop an analytical model whose results fits well to the numerical calculations. From the analytical expressions, a simple protocol for the pulse to be applied on the system in order to reach a given precision is derived. 
Finally, we studied the robustness of the protocol under decoherence effects due to spin-orbit and defects on the chain couplings. The results are weakly dependent on both mechanisms for small and moderate magnitudes of this perturbations. Also in these cases, the analytical model provides an insight on the errors introduced by the decoherence and predicts a transfer of higher accuracy under cubic crossing.
\section{Acknowledgements}
We acknowledge SGCyT-UNNE PI 20T001 and CONICET (PUE2017-22920170100089CO) and (PIP11220200100170CO) for partial financial support.
%------------------

\end{document}